\documentclass[english,aps,preprint]{revtex4}
\usepackage[T1]{fontenc}
\usepackage[latin9]{inputenc}
\usepackage{amsmath}
\usepackage{amssymb}
\usepackage{graphicx}
\usepackage{esint}

\makeatletter
\@ifundefined{textcolor}{}
{%
 \definecolor{BLACK}{gray}{0}
 \definecolor{WHITE}{gray}{1}
 \definecolor{RED}{rgb}{1,0,0}
 \definecolor{GREEN}{rgb}{0,1,0}
 \definecolor{BLUE}{rgb}{0,0,1}
 \definecolor{CYAN}{cmyk}{1,0,0,0}
 \definecolor{MAGENTA}{cmyk}{0,1,0,0}
 \definecolor{YELLOW}{cmyk}{0,0,1,0}
}

\makeatother

\usepackage{babel}
\begin{document}

\preprint{BROWN-HET-1644, NORDITA-2013-36}

\title{Pure states and black hole complementarity}

\author{David A. Lowe}

\email{lowe@brown.edu}

\affiliation{Physics Department, Box 1843, Brown University, Providence, RI, 02912,
USA}

\author{Larus Thorlacius}

\email{larus@nordita.org}

\affiliation{Nordita, KTH Royal Institute of Technology and Stockholm University,
Roslagstullsbacken 23, SE-106 91 Stockholm, Sweden\\{\rm and}\\ University
of Iceland, Science Institute, Dunhaga 3, IS-107 Reykjavik, Iceland}
\begin{abstract}
The future apparent horizon of a black hole develops large stress
energy due to quantum effects, unless the outgoing modes are in a
thermal density matrix at the local Hawking temperature. It is shown
for generic pure states that the deviation from thermality is so small
that an infalling observer will see no drama on their way to the stretched
horizon, providing a derivation of black hole complementarity after
the Page time. Atypical pure states, and atypical observers, may of
course see surprises, but that is not surprising.
\end{abstract}
\maketitle

\section{introduction}

In the usual analysis of Hawking radiation, the problem is analyzed
in either the Hartle-Hawking vacuum \cite{Hartle:1976tp} or the Unruh
vacuum \cite{Unruh:1976db}. The former case is appropriate for eternal
black holes, supported by a thermal flux of radiation from past infinity.
The latter case is not supported by such a flux, and better models
the evaporation of a black hole formed in a dynamical process. Here
back-reaction effects are expected to modify the spectrum of the Hawking
radiation after a substantial fraction of the initial mass is lost.
Each of these vacua involve exterior modes entangled with interior
modes, as described in \cite{Unruh:1976db}. Correlation functions
of local operators outside the black hole horizon may therefore be
viewed as expectation values in a density matrix where at least the
outgoing modes in the infinite future are in a finely tuned thermal
density matrix. The most obvious attempt to modify this situation
by placing such modes in their asymptotic vacuum leads to the Boulware
vacuum \cite{Boulware:1974dm}, which produces a singular renormalized
stress energy tensor on the future horizon (as well as the past horizon).
This leads to violations of the equivalence principle for infalling
observers.

In the present work our goal is to study in more detail deformations
of the Unruh and Hartle-Hawking vacuum states to test the robustness
of the principle of black hole complementarity \cite{Susskind:1993if}.
Generic deformations, in particular those toward pure states, lead
to time dependent fluctuations in the radiation. Such fluctuations
typically lead to divergent energy densities for a freely falling
observer on either the past horizon, the future horizon, or both. 

We begin by establishing that the so-called in-modes may be taken
to be in an arbitrary pure state tensored with either the Hartle-Hawking
or Unruh vacua. This provides us with multi-parameter deformations
of these vacua with finite stress energy on the future horizon. This
provides evidence in favor of the black hole complementarity hypothesis,
by giving examples of smooth deformations of the vacuum states. Moreover,
each of these in-modes has a component outgoing at future infinity,
caused by scattering off the gravitational potential. They provide
examples of outgoing fluxes which do not lead to firewalls on or near
the horizon.

This situation of course is not satisfactory, because these in-modes
may be traced back to the selection of a non-vacuum state at past
infinity. Moreover, we find that any attempt to treat similar excitations
of the out-modes in the Schwarzschild background does indeed lead
to divergent energies as seen by a freely falling observer near the
horizon. Thus, at first sight, it seems out-modes must be locked into
a purely thermal density matrix to avoid drama for an infalling observer.
A finite perturbation at any frequency leads to infinite local energy
densities for infalling observers at the event horizon.

The above discussion refers to a calculation that neglects back-reaction
of the emitted radiation on the geometry. To improve on this situation,
we model the effect of back-reaction using the outgoing Vaidya metric
\cite{Vaidya:1951zz}. This provides a fully time-dependent metric
when a null fluid is emitted from a black hole. We still find a freely
falling observer sees a UV divergent energy density as they approach
the stretched horizon, even with back-reaction included.

Lloyd \cite{Lloyd:1987gy} has pointed out that random pure states
can lead to effects that mimic averaging over ensembles in statistical
mechanics (see also later work by Page \cite{Page:1993df} where it
was emphasized information does not begin to emerge from a black hole
until a time of order $M^{3}$, which we refer to as the Page time).
In fact, there is a sense in which the convergence is much more rapid.
If a reduced density matrix is constructed by tracing over a Hilbert
subspace of dimension $e^{N}$ the error in the density matrix is
of order $e^{-N/2}$. If one instead computes fluctuations in a statistical
ensemble, the finite size effects are typically much larger, of order
$1/\sqrt{N}$. We use this observation to show that an infalling observer
into a generic pure state black hole will not see any drama up to
the stretched horizon. If the black hole is projected into an outgoing
mode eigenstate the infaller can indeed see mild drama as they approach
the stretched horizon as noted in the previous paragraph, but projections
are either non-generic or impractical. These results provide a derivation
of black hole complementarity for black holes older than the Page
time.

It should be emphasized that we take care to use local unitary effective
field theory only outside the stretched horizon, where it has a conventional
interpretation. One may also try to build effective field theory on
patches of spacetime inside the horizon, however the interpretation
there is much more problematic. Inside the horizon physical observables
are inherently imprecise, and there is much room to hide highly non-local
physics \cite{Lowe:2006xm}. An exact fundamental description may
predict non-local physics inside the horizon that is not captured
by an approximate local unitary effective field theory in that region.
Conversely, applying local effective field theory across the horizon
will predict effects that are not realized in the fundamental unitary
description.

\section{Non-rotating black hole evaporation in 3+1 dimensions: Problems and
solutions}

\subsection{Mode expansions and vacua}

In this section we consider a massless conformally coupled scalar
field. Issues of back-reaction will be ignored, and re-examined in
the following section. The metric in Schwarzschild coordinates takes
the form
\[
ds^{2}=-\left(1-\frac{2M}{r}\right)dt^{2}+\frac{dr^{2}}{1-\frac{2M}{r}}+r^{2}d\theta^{2}+r^{2}\sin^{2}\theta d\phi^{2}\,.
\]
In these coordinates, a complete set of modes in the exterior region
may be obtained by separating the equation of motion, and defining
the tortoise radial coordinate
\[
r_{*}=r+2M\log\left(\frac{r}{2M}-1\right)\,.
\]
The angular and time dependence may be handled straightforwardly,
and the radial equation can be mapped into a scattering problem with
a step-like potential separating the behavior at $r\to\infty$ from
the region $r\to2M$ \cite{DeWitt:1975ys}. This leads to a natural
decomposition into independent modes that we refer to as in-going
and out-going %
\footnote{Following \cite{Christensen:1977jc}, we correct a typo that appears
in \cite{DeWitt:1975ys}.%
}:
\begin{eqnarray}
u^{in}(x) & = & \left(4\pi\omega\right)^{-1/2}e^{-i\omega t}R_{l}^{in}(\omega;r)Y_{lm}(\theta,\phi)\nonumber \\
u^{out}(x) & = & \left(4\pi\omega\right)^{-1/2}e^{-i\omega t}R_{l}^{out}(\omega;r)Y_{lm}(\theta,\phi)\label{eq:modes}
\end{eqnarray}
with
\begin{eqnarray*}
R_{l}^{out}(\omega;r) & \sim & \begin{cases}
r^{-1}e^{i\omega r_{*}}+A_{l}^{out}(\omega)r^{-1}e^{-i\omega r_{*}}, & r\to2M\\
B_{l}(\omega)r^{-1}e^{i\omega r_{*}}, & r\to\infty
\end{cases}\\
R_{l}^{in}(\omega;r) & \sim & \begin{cases}
B_{l}(\omega)r^{-1}e^{-i\omega r_{*}}, & r\to2M\\
r^{-1}e^{-i\omega r_{*}}+A_{l}^{in}(\omega)r^{-1}e^{i\omega r_{*}}, & r\to\infty\,.
\end{cases}
\end{eqnarray*}
Scattering off the gravitational field leads to ``grey body'' factors,
so a mode that is purely outgoing near infinity contains an ingoing
component near the horizon, and likewise a mode that is purely ingoing
near the horizon contains an outgoing component near infinity. 

The Unruh vacuum is defined by requiring the modes incoming at past
null infinity to be purely positive frequency with respect to $t$,
and while those outgoing from the past horizon are positive frequency
with respect to the appropriate Kruskal coordinate. This vacuum corresponds
to an evaporating black hole with no incoming flux at past infinity,
but a thermal outgoing flux at future null infinity.

The Hartle-Hawking vacuum is defined in a similar way, except the
condition at past null infinity is replaced by the condition that
infalling modes on the future horizon are positive frequency with
respect the appropriate Kruskal coordinate. This corresponds to an
eternal black hole with balancing ingoing and outgoing thermal fluxes
at infinity.

It is also worth mentioning the Boulware vacuum where $t$ is used
to define positive frequency throughout the exterior region. This
vacuum leads to singular quantum corrections at the horizon.

\subsection{Fluctuations\label{sub:Fluctuations}}

In the following we will mostly be interested in the Unruh vacuum,
which describes an evaporating black hole. An important set of early
results in this direction was developing an understanding of renormalization
in this curved background \cite{Christensen:1977jc} which led to
explicit computations of the one-loop corrections to $\left\langle \phi^{2}\right\rangle $
and $\left\langle T_{\mu\nu}\right\rangle $ for a massless scalar
field in the Schwarzschild background, in the Unruh, Hartle-Hawking
and Boulware vacua \cite{Candelas:1980zt}. In both the Unruh and
Hartle-Hawking vacua, these corrections were found to be mild, leading
to the expectation that back-reaction near the horizon should be negligible.
We will discuss these computations in more detail in the following
subsection.

\begin{figure}

\includegraphics{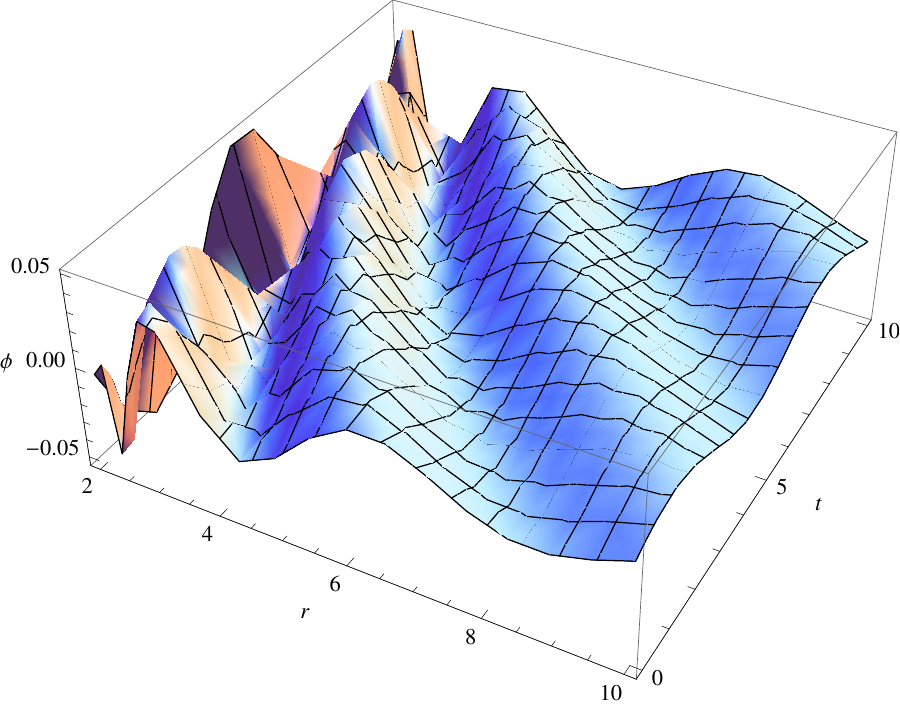}\caption{\label{fig:Scalar-mode-fluctuation}Scalar mode fluctuation near the
horizon. Here we set $M=1$ and $\omega=1$.}
\end{figure}

For the moment, let us study the behavior of the individual modes
\eqref{eq:modes} near the past and future horizons. For both the
ingoing and outgoing modes, the mode functions are finite as $r\to2M$
but oscillate more and more rapidly with $r$ as the horizon is approached.
Fixed frequency oscillations appear with respect to the time coordinate
$t$. This is illustrated in figure \ref{fig:Scalar-mode-fluctuation}.

\begin{figure}
\includegraphics{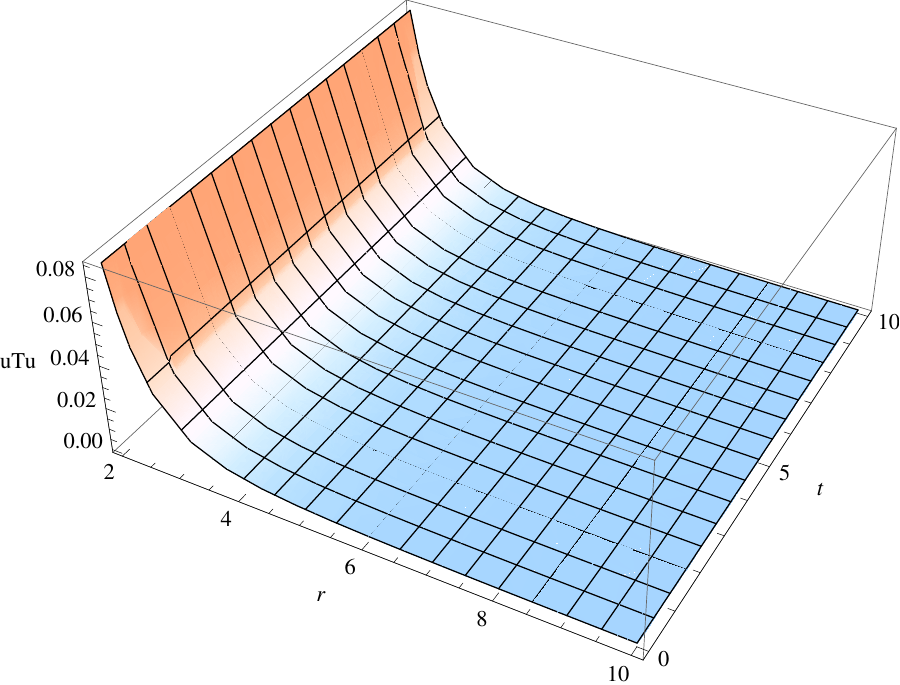}

\caption{\label{fig:The-expectation-value}The expectation value of an ingoing
fluctuation contracted with the velocity of an infalling timelike
geodesic near the future horizon. Here we set $M=1$ and $\omega=1$.}

\end{figure}
The stress energy tensor for a massless conformally coupled scalar
field in a Ricci flat background is
\[
T_{\mu\nu}=\frac{2}{3}\partial_{\mu}\phi\partial_{\nu}\phi-\frac{1}{6}g_{\mu\nu}\partial_{\rho}\phi\partial^{\rho}\phi-\frac{1}{3}\phi_{;\mu\nu}\phi+\frac{1}{12}g_{\mu\nu}\phi\square\phi\,.
\]
To study the behavior of this quantity near the horizon, we must first
contract indices with some suitably defined basis vectors. Near the
future horizon, we choose a velocity 4-vector corresponding to a time-like
radial ingoing geodesic 
\begin{equation}
u^{\mu}=\left(\frac{k}{1-\frac{2M}{r}},-\sqrt{k^{2}-1+\frac{2M}{r}},0,0\right)\,,\label{eq:velocity}
\end{equation}
in a $(t,r,\theta,\phi)$ basis. The result for an ingoing mode is
shown in figure \ref{fig:The-expectation-value}. The answer is finite
on the horizon, and independent of time. 

If we perform the same computation for an ingoing mode on the past
horizon and instead choose an outgoing radial timelike geodesic, we
find a double pole as $r\to2M$ and a divergent result. The result
is again independent of time. 

The outgoing modes produce a stress tensor that is singular on both
the past and future horizons, with rapid oscillations combined with
double pole terms as $r\to2M.$ Now the stretched horizon is placed
at a value of $r$ such that the red-shift to infinity is a constant,
such that 
\begin{equation}
\Lambda_{UV}=\frac{M_{pl}^{2}}{M}\frac{1}{\sqrt{1-\frac{2M}{r}}}\,,\label{eq:redshift}
\end{equation}
where $\Lambda_{UV}$ will be the ultraviolet cutoff scale for the
stretched horizon theory, which may be taken to be some energy below
the Planck scale. The double pole indicates the infalling observer
sees a large energy density
\begin{equation}
\rho\sim T^{4}\frac{1}{\left(1-\frac{2M}{r}\right)^{2}}=\Lambda_{UV}^{4}\,.\label{eq:enden}
\end{equation}

As we will see in the next subsection, if we sum over modes to compute
the correct one-loop contributions to the vacuum expectation values
of these quantities, and correctly renormalize \cite{Christensen:1977jc,Candelas:1980zt},
there are delicate cancellations that remove the future horizon divergence,
in the case of the Unruh vacuum; and for both horizons in the case
of the Hartle-Hawking vacuum. It will then be our goal to model time-dependent
pure state corrections to these results.

\subsection{Correlators}

Let us begin by studying the simplest quantity built out of the scalar
field that receives quantum corrections and can become potentially
divergent on the horizon $\left\langle \phi^{2}\right\rangle $. As
we saw in the previous subsection, the modes themselves are finite
on the horizon, but derivative operators such as $T_{\mu\nu}$ may
become singular. Following \cite{Candelas:1980zt} we can construct
$\left\langle \phi^{2}\right\rangle $ by applying a point-splitting
regularization to the tree-level propagator in the appropriate vacuum
state, then applying a local counter-term subtraction procedure.

For the Unruh vacuum $|U\rangle$, this yields
\begin{eqnarray}
\left\langle U|\phi^{2}|U\right\rangle  & = & \frac{1}{16\pi^{2}}\int_{0}^{\infty}\frac{d\omega}{\omega}\left[\sum_{l=0}^{\infty}(2l+1)\left(\coth\frac{\pi\omega}{\kappa}\left|R_{l}^{out}(\omega;r)\right|^{2}+\left|R_{l}^{in}(\omega;r)\right|^{2}\right)-\frac{4\omega^{2}}{1-\frac{2M}{r}}\right]\nonumber \\
 &  & -\frac{4M^{2}}{48\pi^{2}r^{4}\left(1-\frac{2M}{r}\right)}\,,\label{eq:phisq}
\end{eqnarray}
where $\kappa=\frac{1}{4M}$ is the surface gravity at the horizon.
The first term corresponds to the outgoing modes, the second the ingoing
modes, and the last two terms correspond to the counter-term contributions.
The appearance of the $\coth\frac{\pi\omega}{\kappa}$ factor is a
consequence of the thermality of the outgoing modes. In the Hartle-Hawking
vacuum, such a factor also appears in front of the ingoing term. 

The sum over angular momenta of the outgoing term yields only a partial
cancellation of the $r\to2M$ pole, while the sum for the ingoing
term is finite in this limit. Only after integrating over frequency
is the $r\to2M$ pole canceled. This requires a delicate exact cancellation
between the counter-terms and the thermal outgoing modes.

It is worth noting any finite excitation of the Unruh vacuum by ingoing
modes preserves the finiteness of $\left\langle \phi^{2}\right\rangle $.
Thus there is an easily accessible collection of modifications of
the Unruh vacuum obtained by tensoring in essentially arbitrary infalling
pure states that leads to finite stress energy near the horizon.

However to have a successful theory of the stretched horizon, this
is necessary, but not sufficient. If the Unruh vacuum is to be replaced
by a pure state built out of stretched horizon modes, and exterior
modes, and the Hawking radiation is to be produced by unitary evolution,
then the stretched horizon must also be capable of emitting outgoing
modes in a manner that deviates from exact thermality. We turn to
this question in the next subsection and examine whether back-reaction
ameliorates the problem.

\subsection{Outgoing Vaidya metric\label{sub:Outgoing-Vaidya-metric}}

\begin{figure}
\includegraphics[clip,scale=0.75]{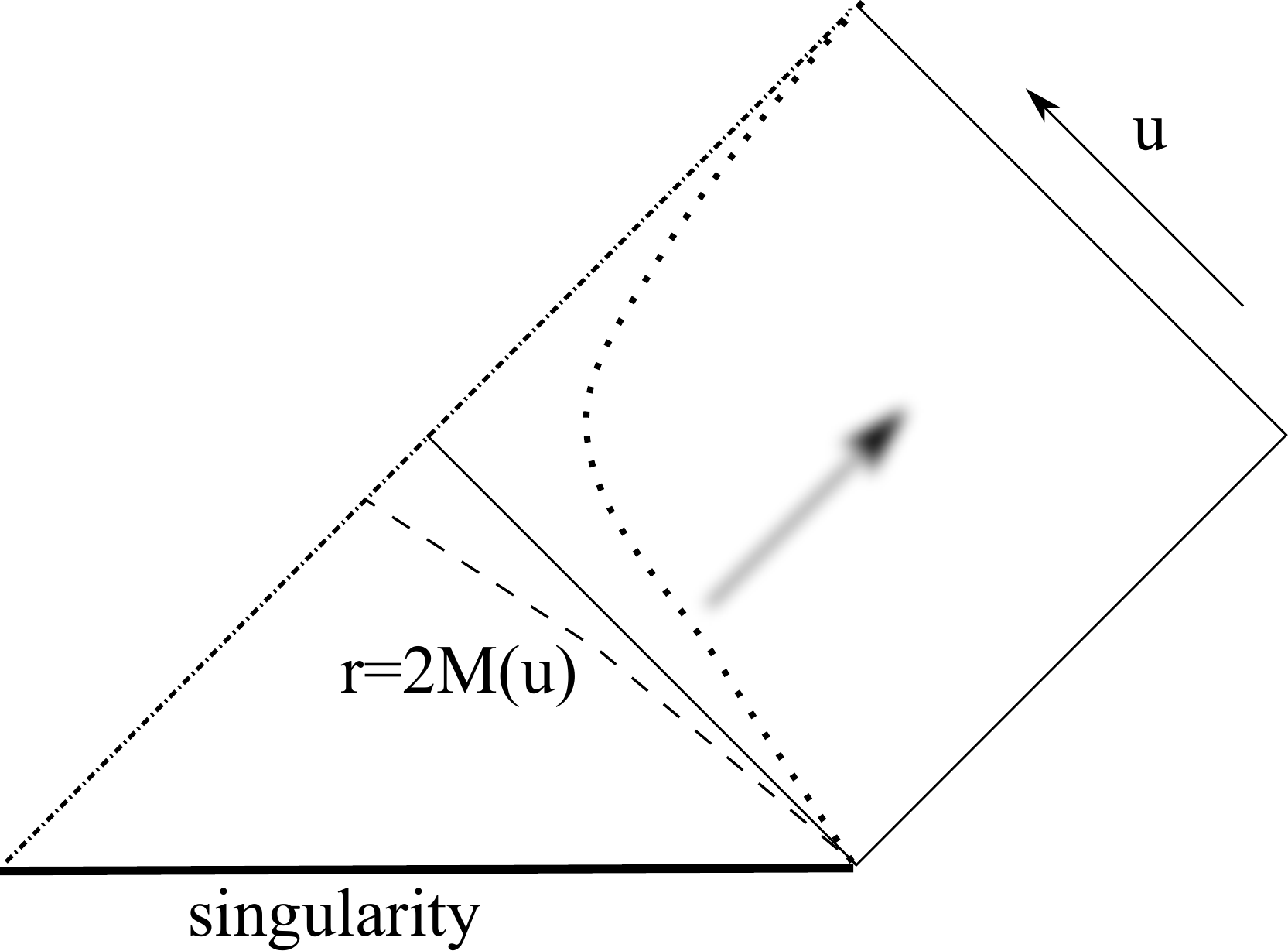}

\caption{The Penrose diagram for the outgoing Vaidya metric. The lower dashed
line shows the apparent horizon below the global horizon. The upper
dot-dashed line indicates where the metric is geodesically incomplete,
and may be patched onto a variety of interior solutions. The dotted
timelike line shows the position of the stretched horizon. \label{fig:The-Penrose-diagram}}

\end{figure}

In the above we have seen that a single classical outgoing mode of
definite frequency induces an infinite stress energy on the global
horizon after taking into account the effects of renormalization.
Let us now see if this divergence survives if we also include gravitational
back-reaction. This kind of problem has been studied extensively in
the literature, for example in the study of neutrino emission during
stellar core collapse \cite{Misner:1965zza,Lindquist:1965zz} in the
limit of spherical symmetry. The emission of massless matter, in a
so-called null fluid, may be studied analytically using the outgoing
Vaidya metric \cite{Vaidya:1951zz}
\[
ds^{2}=-\left(1-\frac{2M(u)}{r}\right)du^{2}-2dudr+r^{2}d\Omega^{2}\,,
\]
with stress energy tensor
\[
T^{\mu\nu}=-\frac{1}{4\pi r^{2}}\frac{dM}{du}k^{\mu}k^{\nu}\,,
\]
where $k^{\mu}$ is a null-vector directed radially outward, normalized
as in \cite{Lindquist:1965zz}. Some properties of the solution are
illustrated in figure \ref{fig:The-Penrose-diagram}.

An infalling observer will measure an energy density in her reference
frame 
\[
\rho_{in}=T^{\mu\nu}U_{\mu}U_{\nu}\,,
\]
where $U^{\mu}$ is the velocity 4-vector of the infalling observer.
This is obtained by solving the equation for a timelike geodesic in
these coordinates. For a purely radial motion, the components of the
velocity are
\[
U^{\mu}=\left(\frac{du}{d\tau},\frac{dr}{d\tau},\frac{d\theta}{d\tau},\frac{d\phi}{d\tau}\right)=\left(\frac{1}{V+\sqrt{V^{2}+1-\frac{2M(u)}{r}}},V,0,0\right)\,.
\]
The energy density of the infaller is then \cite{Lindquist:1965zz}
\begin{equation}
\rho_{in}=-\frac{1}{4\pi r^{2}}\frac{dM}{du}\frac{1}{\left(V+\sqrt{V^{2}+1-\frac{2M(u)}{r}}\right)^{2}}\,.\label{eq:infallenergy}
\end{equation}
To model the process of interest to us, let us consider the solution
corresponding to the emission of energy $M_{pl}^{2}/M$ over a time
$M/M_{pl}^{2}$ measured at infinity. Inserting these values into
\eqref{eq:infallenergy}, and taking $V<0$ corresponding to an infalling
observer, we find that near the apparent horizon the energy density
becomes
\begin{equation}
\rho_{in}=\frac{1}{4\pi}\frac{M_{pl}^{8}}{M^{4}}\frac{V^{2}}{(1-\frac{2M(u)}{r})^{2}}\,.\label{eq:infallen2}
\end{equation}
Using formula \eqref{eq:redshift} this expression may be rewritten
\begin{equation}
\rho_{in}=\frac{V^{2}}{4\pi}\,\Lambda_{UV}^{4}\,.\label{eq:enestimate}
\end{equation}
This shows that the UV-divergence persists when the gravitational
back-reaction due to the outflow of energy is taken into account.
This provides strong evidence that even with back-reaction included,
any time dependence of the outgoing radiation will lead an infaller
to effectively see a firewall as they approach the stretched horizon.

\subsection{Near-horizon observables}

We saw in the previous subsections, that while we are free to modify
the infalling modes at will, delicate cancellations are needed with
the outgoing modes to yield a finite renormalized $\left\langle \phi^{2}\right\rangle $
at the horizon. Obtaining finite stress energy involves even further
cancellations. As we saw in section \ref{sub:Fluctuations} even a
single outgoing mode generates singular contributions to the stress
energy. 

To formulate these issues more sharply, let us consider a freely falling
infalling observer, who is capable of measuring with a UV cutoff $\Lambda_{UV}$.
We model the stretched horizon theory by a surface that emits quanta
of energy $M_{pl}^{2}/M$ every $M$ units of time $t$. If the infaller
falls in after a finite fraction of the black hole lifetime $M^{3}$,
the infaller crosses a substantial fraction of the outgoing Hawking
radiation, of order $\left(\frac{M}{M_{pl}}\right)^{2}$ particles.
Since a freely falling observer will hit the singularity in proper
time less than of order $M$, the infaller sees $\frac{M}{M_{pl}}$
outgoing modes per unit Planck time. Therefore a freely falling observer
cannot resolve individual Hawking modes, and her local operators will
involve linear combinations of at least $M/M_{pl}$ outgoing modes.
If the infaller makes one local measurement every time $1/\Lambda_{UV}$
in her rest frame, then the subspace of the Hilbert space accessible
in her lifetime $M$ will have dimension $n_{in}=e^{M/\Lambda_{UV}}\ll e^{M/M_{pl}}$. 

Let us denote the subspace of the Hilbert space corresponding to this
outgoing radiation $A$, and the subspace corresponding to the stretched
horizon degrees of freedom $B$. The rest of the outgoing radiation
we denote by the Hilbert subspace $C$. According to the postulates
of black hole complementarity, the combined Hilbert space $A\times B\times C$
undergoes local unitary evolution, mapping a pure state to a pure
state. We expect the dimension of $A$ will have an upper bound of
order $n_{in}$ while the dimension of $B$ and $C$ will be of order
$e^{M^{2}/M_{pl}^{2}}$, assuming we are not too close to the endpoint
of evaporation %
\footnote{In these formulas for large numbers, such as $e^{M^{2}/M_{pl}^{2}}$
, of order this number should be taken to mean of order $e^{cM^{2}/M_{pl}^{2}}$
where $c$ is a finite constant.%
}. As time passes the dimensions of these Hilbert subspaces will shift,
but the combined dimension will remain constant.

We wish to compute the expectation value of the stress energy tensor
seen by an infalling observer in a generic pure state emitted after
scrambling on the stretched horizon. As we have seen above, the infalling
modes may be placed in an arbitrary pure state, leading to finite
corrections to the expectation value. We therefore focus our attention
on the contribution due to the outgoing modes.

Now fluctuations in the stress energy tensor can only become large
in the limit that $r\to2M(u)$, which follows from the $r\to2M$ divergent
terms in \eqref{eq:phisq} as shown in \cite{Candelas:1980zt}. The
modes relevant for determining whether the infaller sees a large effect
are those emitted within $\delta u\sim M(u)$, so even though these
modes free stream from the stretched horizon, they were in relatively
recent causal contact with the stretched horizon degrees of freedom.
This implies in this period of time the $A\times B$ subsystem evolves
unitarily on its own, so that $\delta S_{A}=-\delta S_{B}$ and $\delta E_{A}=-\delta E_{B}$.
Thus, the effective temperatures of these systems are the same
\[
\frac{1}{T}=\frac{\partial S_{A}}{\partial E_{A}}=\frac{\partial S_{B}}{\partial E_{B}}\,.
\]
However, to within small corrections in the temperature, one can consider
a time period just before the emission of $A$ from the stretched
horizon, and likewise argue that
\[
\frac{1}{T}=\frac{\partial S_{C}}{\partial E_{C}}=\frac{\partial S_{B}}{\partial E_{B}}\,.
\]
Thus all subsystems are at the same effective temperature $T=M_{pl}^{2}/M(u)$
to within negligible corrections, so the evolution on $A\times B\times C$
may be treated in an adiabatic approximation. 

In quantum field theory, local operators are constructed to model
the action of real detectors, and likewise local operators may be
used to prepare initial states of interest. The resulting correlators
of the local fields may be interpreted as probability amplitudes and
used to predict the outcomes of experiments. As is typical in quantum
mechanics, the outcome of a particular measurement is determined probabilistically,
which effectively leads to a version of averaging that mimics the
averaging in statistical mechanics \cite{Lloyd:1987gy}. One of the
key points of that work is that expectation values in a random pure
state converge much more rapidly than the ensemble averages used in
ordinary statistical mechanics. It was found that fluctuations in
an expectation value are typically suppressed by a factor $1/\sqrt{n}$
where $n$ is the dimension of the Hilbert subspace that is averaged
over in selecting a random pure state. This comes from the integrating
over a shell in the space $\mathbb{C}^{n}$. This is to be contrasted
with the usual suppression of fluctuation from ensemble averages which
are of the order $1/\sqrt{N}$ where $N$ is the number of degrees
of freedom in the system averaged over (typically $n\sim e^{N}$). 

Unfortunately it is difficult to make these ideas precise in a completely
general context. For example a pure state which is an eigenstate of
some particular operator that commutes with the Hamiltonian will remain
in that eigenstate for all time, and any effective measurements that
commute with this operator will only produce that eigenvalue. This
makes the definition of a complex pure state a rather basis dependent
question.

However we can make these statements rather more precise in the context
of measurements of the evaporation of a black hole. The natural basis
for an observer far from the black hole is indeed the outgoing modes
discussed above. However such modes are highly unnatural from the
viewpoint of a freely falling observer near the horizon. 

Applying this to the case at hand, any operator corresponding to the
detector of a freely infalling observer will average over the subspace
$B\times C$. Since the operator is local, it will not probe the subspaces
$B$ and $C$. One may therefore compute the expectation value by
tracing over the Hilbert subspaces $B$ and $C$ to produce the reduced
density matrix $\rho_{A}$. At late times, the modes $A$ will be
maximally entangled with the earlier radiation $C$ \cite{Hayden:2007cs}.
By the arguments of \cite{Lloyd:1987gy} this density matrix will
agree with the canonical ensemble at temperature $T$ up to corrections
of order $e^{-S_{C}/2}$. 

Let us briefly review this computation in more detail \cite{Lloyd:1987gy}.
Let us assume we have a pure state on a product Hilbert space $A\times C$
described by the density matrix $\rho_{AC}$ with total energy $E$.
Defining
\[
\rho_{A}=\mathrm{Tr}_{C}\rho_{AC}=\sum_{i}p(E_{i})\rho_{E_{i}}\,.
\]
Here $E_{i}$ are the energy eigenvalues of the subspace $A$, $i$
labels the energy eigenstates, $p(E_{i})$ is the probability of occupation
of energy eigenstate $E_{i}$ and $\rho_{E_{i}}$ is a density matrix
on the subspace of $A$ corresponding to eigenstates with energy $E_{i}$.
Computing this for a typical pure state, one finds
\[
\rho_{A}=\frac{1}{n}\sum_{i}e^{S_{A}(E_{i})+S_{C}(E-E_{i})}\rho_{E_{i}}\left(1\pm\mathcal{O}\left(\frac{1}{e^{S_{C}(E-E_{i})/2}}\right)\right)\,,
\]
where we have assumed $S_{C}\gg S_{A}$, and defined $n$ as the total
dimension of the Hilbert space $A\times C$. The exponential may be
approximated using $S_{C}(E-E_{i})\approx S_{C}(E)-E_{i}/T$, using
$1/T=\partial S_{C}/\partial E$, leading to the canonical ensemble
expression, up to small corrections
\[
\rho_{A}=\frac{1}{N}\sum_{i}e^{-E_{i}/T+S_{A}(E_{i})}\rho_{E_{i}}\left(1\pm\mathcal{O}\left(\frac{1}{e^{S_{C}(E)/2}}\right)\right)\,,
\]
with $N=e^{S_{C}(E)}/n$.

This density matrix may then be used to estimate 
\begin{equation}
\left\langle T_{\mu\nu}\right\rangle U^{\mu}U^{\nu}\sim\frac{e^{-S_{C}/2}}{\left(1-\frac{2M(u)}{r}\right)^{2}}\,,\label{eq:vevstress}
\end{equation}
as $r\to2M(u)$ by viewing the correction as a classical contribution
to the emitted energy in the outgoing Vaidya solution \eqref{eq:infallen2}.
While this still becomes singular very close to the global horizon,
this is safely behind the stretched horizon, and in that region we
do not trust conventional effective field theory. We conclude an infalling
observer sees no drama in their approach to the stretched horizon
for a generic pure state.

\subsection{EPR paradox in the black hole setting}

The above argument suggests that the infalling observer sees smooth
stress energy all the way up to the stretched horizon, beyond which
it is difficult to make model-independent statements. However we run
into an apparent paradox if we suppose that an external observer far
from the black hole projects it onto an eigenstate of the outgoing
modes. In this case, the model of section \ref{sub:Outgoing-Vaidya-metric}
should provide an accurate estimate of the stress energy, and we expect
the infalling observer to see a firewall. 

The resolution is very similar to that of the original EPR paradox.
Suppose the infalling observer is initially spacelike separated from
the outside observer. His measurements are unaffected by the outside
observer's measurements. But nevertheless the measurements can be
correlated via the nonlocality of ordinary quantum mechanics. Effectively
the density matrix $\rho_{A}$ corresponds to a trace over macroscopic
superpositions of the states of the outside observer. Only in a generic
superposition is the correlator $\left\langle T_{\mu\nu}\right\rangle U^{\mu}U^{\nu}$
finite. 

A measurement in the state projected by the outside observer $\left\langle T_{\mu\nu}\mathcal{O}_{M^{2}}\right\rangle U^{\mu}U^{\nu}$
is expected to be large, where the operator $\mathcal{O}_{M^{2}}$
represents the measurement of the outside observer on $\frac{M^{2}}{M_{pl}^{2}}$
Hawking particles. However the unnaturally large value for this correlator
only appears as a puzzle to the outside observer if he is able to
accelerate away from the black hole and compare notes with the distant
observer. It has no local significance to the infaller, except in
the atypical situation when the black hole is prepared in such an
eigenstate from the beginning. However here we may rely on the fact
that the likelihood of such an eigenstate of of order $e^{-M^{2}/M_{pl}^{2}}$.

Another variant on this process involves an observer who stays outside
the black hole for a long time to precisely measure its state, and
then falls in. Perhaps not surprisingly, such an observer can predict
the emission of non-thermal Hawking particles and choose to fall into
the horizon to measure them. Such an observer will similarly see a
large effect of order \eqref{eq:enestimate} near the stretched horizon.
However the practicality of these measurements seems unlikely. Such
an observer would need energy and entropy with which to store all
this data, comparable to those of the black hole he is reconstructing.
This process would be well-approximated by the collision of two black
holes of similar mass. In such a collision, Planck-scale curvatures
are not produced in the vicinity of the apparent horizon(s), but there
is nevertheless a substantial fraction of the initial Bondi energy
radiated in terms of gravitational radiation. It is interesting to
note that gravitational effects show a tendency to smooth out would
be curvature/stress energy singularities. We conclude that just as
atypical pure states can give surprising answers, we may also have
atypical observers who are surprised by their measurements. 

Finally, one can try to imagine a single Hawking particle plays the
role of the observer, to parallel the arguments of \cite{Almheiri:2012rt},
who instead conclude a firewall exists at the horizon. Related arguments
have been made in \cite{Mathur:2011uj,Mathur:2012jk} in the context
of the fuzzball scenario. The arguments made in these works have already
been rebutted in \cite{LarjoLowe} and in the present work extend
and strengthen this approach. In the case of a single Hawking particle
the ``observer'' only has access to a 1-dimensional subspace of
the Hilbert space, so once again it is appropriate to trace over the
other subspaces. For an infalling Hawking particle, we reproduce \eqref{eq:vevstress}
with the velocity $U^{\mu}$ replaced by a normalized null vector.
For an outgoing Hawking particle we get a negligible result. We conclude
therefore that neither Hawking particles, nor infalling observers
see drama near the stretched horizon of a black hole in a generic
pure state. However it is possible to choose a special pure state,
or even a special observer where this conclusion does not hold.

\section{Discussion}

To extend these considerations to an observer falling across the horizon,
one would need to account for the fact that the mapping from the fundamental
unitary description to the effective description is no longer local.
The rules of unitary and locality in the bulk must then be given up.
Some early work which found that local effective field theory does
not predict its own demise when horizons are present appeared in \cite{Lowe:1995ac,Lowe:1995pu}.
Rather we expect local unitary effective field theories %
\footnote{Note that nonlocal effective theories cannot be quantized in general.%
} are capable of approximately describing the measurements that may
be carried out by an infalling observer. However these will disagree
with the exact answers of a unitary nonlocal holographic description
of the same measurements \cite{Lowe:2006xm}. Related ideas have been
considered more recently in the context of the firewall scenario using
a quantum computational model in \cite{Harlow:2013tf}. Evidence for
such a scenario has been provided using the AdS/CFT framework in \cite{Lowe:2009mq}.
This scenario has a chance of working, because the finite lifetime
of an infalling observer limits the measurement operations that may
be carried out, thus the effective field theory in a region inside
the horizon need not give exact answers.

From the viewpoint of evolution of the stretched horizon theory, an
infalling observer's degrees of freedom evolve for a time of order
the scrambling time, before being reemitted in the Hawking radiation.
The scrambling time, measured at infinity, thus provides a time-scale
at which the evolution of these degrees of freedom qualitatively changes.
It is tempting to match this delay time with the proper time that
the infaller takes to hit the singularity. We hope to return to this
question using more specific models of the stretched horizon theory
in future work, though progress has already been made \cite{Lowe:2006xm,Lowe:2009mq}.
\begin{acknowledgments}
D.L. thanks Daniel Harlow, Daniel Jafferis and Savvas Koushiappas
for helpful discussions. The research of D.A.L. is supported in part
by DOE grant DE-FG02-91ER40688-Task A and an FQXi grant. The research
of L.T. is supported in part by grants from the Icelandic Research
Fund and the University of Iceland Research Fund.
\end{acknowledgments}
\bibliographystyle{apsrev}
\bibliography{firewall}

\end{document}